\documentstyle[twocolumn,prl,epsf,aps]{revtex}
\newcommand{\nix}[1]{}

\begin{document}

\title{Circular photogalvanic effect induced by monopolar spin
orientation in $p$-GaAs/AlGaAs MQW}

\author{S.D.~Ganichev$^{1,2}$, E.L.~Ivchenko$^2$,
H.~Ketterl$^1$,  L.E.~Vorobjev$^3$ and W.~Prettl$^1$}

\address{$^1$ Institut f\"ur Experimentelle und Angewandte Physik,
Universit\"at Regensburg\\ 93040~Regensburg, Germany \\ $^2$
A.F.~Ioffe Physico-Technical Institute, Russian Academy of Sci.\\ 194021
St.~Petersburg, Russia\\ $^3$ St.
Petersburg State Technical University, 195251 St.
Petersburg, Russia}

\maketitle
\begin{abstract}

The circular photogalvanic effect (CPGE) has been observed in
(100)-oriented $p$-GaAs/AlGaAs quantum wells at normal incidence of
far-infrared  radiation.
It is shown that monopolar optical spin orientation of free carriers
causes an electric current which
reverses its direction upon changing from left to right
circularly
polarized radiation.
CPGE at normal incidence and the occurence of the linear
photogalvanic effect indicate a
reduced point symmetry of studied multi-layered
heterostructures.
As proposed, CPGE can be utilized to investigate
separately spin polarization of electrons and holes and the
symmetry of quantum wells.
\end{abstract}

Spin polarization and spin relaxation in semiconductors are the subject
of intensive studies of spin-polarized electron transport aimed at the
development of spinotronic devices like a spin
transistor~\cite{Datta,Prinz}. So far,  photo-induced spin
polarization has been achieved by {\em interband} optical absorption of
circularly polarized light with the photogeneration of spin-polarized
electrons and holes~\cite{Meier,Awschalom1,Awschalom2}.

Here we report on the first observation of the circular photogalvanic
effect (CPGE) under {\em intersubband} transitions in
multiple-quantum-well (MQW) structures. Phenomenologically, the
effect is a transfer of angular momenta of circularly polarized photons
into the directed movement of free carriers, electrons or holes, and
therefore depends on the symmetry of the medium. Microscopically, it
arises due to optical spin orientation of holes in MQWs and
asymmetric spin-dependent scattering of spin-polarized carriers by
interface imperfections followed by an appearance of an electric
current. The two states of light circular polarization $\sigma_{\pm}$
result in different spin orientations and, thus, in electric
photocurrents of opposite directions. The CPGE has been
observed on $p$-doped (001)-oriented MQWs at normal incidence of the
circularly polarized far-infrared (FIR) laser pulses. In contrast
to the case of {\em interband} optical excitation, under {\em
intersubband} transitions only one kind of carriers is involved
leading to a monopolar spin orientation. The observed effect can be
utilized to investigate spin polarization of extrinsic free carriers,
and provides a simple method to analyze the reduced point symmetry of
the medium under study, here MQWs.

The experiments have been carried out on $p$-GaAs/AlGaAs (001)-MOCVD grown
quantum well (QW) structures with 400 undoped wells of 20~nm width
separated by 10~nm wide doped barriers. Different samples of
5$\times$5\,mm$^2$ size with hole densities in the QWs varying from
$2\cdot10^{11}$\,cm$^{-2}$ to $2\cdot10^{12}$\,cm$^{-2}$
have been investigated. Ohmic contacts to the QWs were formed
by depositing Ti and Au with subsequent tempering. Two pairs
of point contacts along two perpendicular connecting lines have been
prepared, parallel to [1$\bar{1}$0] and [110] (see inset Fig.~1). Two
additional pairs of ohmic contacts have been formed in the
corners of the sample corresponding to the $\langle 100 \rangle$
crystallographic directions.

A high power pulsed far-infrared molecular laser pumped by a TEA-CO$_2$
laser has been used as a radiation source delivering 100\,ns pulses
with the intensity $I$ up to 1\,MW/cm$^2$. NH$_3$ has been used as a
FIR laser medium yielding strong single-line linearly-polarized
emissions at wavelengths $\lambda$ = 76\,$\mu$m, 90\,$\mu$m, and
148\,$\mu$m~\cite{Ganichev97,Ganichev991}. The photon energies of the
laser lines correspond to transitions between size-quantization
subbands of GaAs QWs\cite{Vorobev96}. The linearly-polarized laser
light could be modified to a circularly-polarized radiation by
applying crystalline quartz $\lambda/4$--plates.

While irradiating the (001) GaAs QWs by normally incident
circularly-polarized radiation (see inset in Fig.~1) a fast
photocurrent signal, $U_{12}$, has been observed in unbiased samples
across one contact pair (contacts 1 \& 2 in the inset). The signal
follows the temporal structure of the laser pulse intensity. The
photocurrent changes sign if the circular polarization is
switched from $\sigma_+$ to $\sigma_-$. Measurements of $U_{12}$ as a
function of the degree of circular polarization $P_{circ} =
\sin\,2\varphi$, where $\varphi$ is the angle between the optical
axis of the $\lambda/4$-plate and the plane of polarization
of the laser radiation, reveal that the photogalvanic
current $j$ is proportional to $P_{circ}$ (Fig.~1). In case of the
linearly-polarized radiation corresponding to $\varphi = 0$ or
$90^{\circ}$ the signal $U_{12}$ vanishes. The magnitude and
the sign
of the CPGE are practically unchanged with variation of the
angle of incidence  in the range of -40$^{\circ}$ to
+40$^{\circ}$. Heating the sample from 4.2~K to 300~K leads to
a change of the sign of  CPGE but the $\sin\,2\varphi$
dependence is retained. In bulk GaAs substrate material no
CPGE could be detected which gives evidence that the effect
occurs in the MQWs.
\begin{figure}
   \centerline{\epsfxsize 10cm \epsfbox{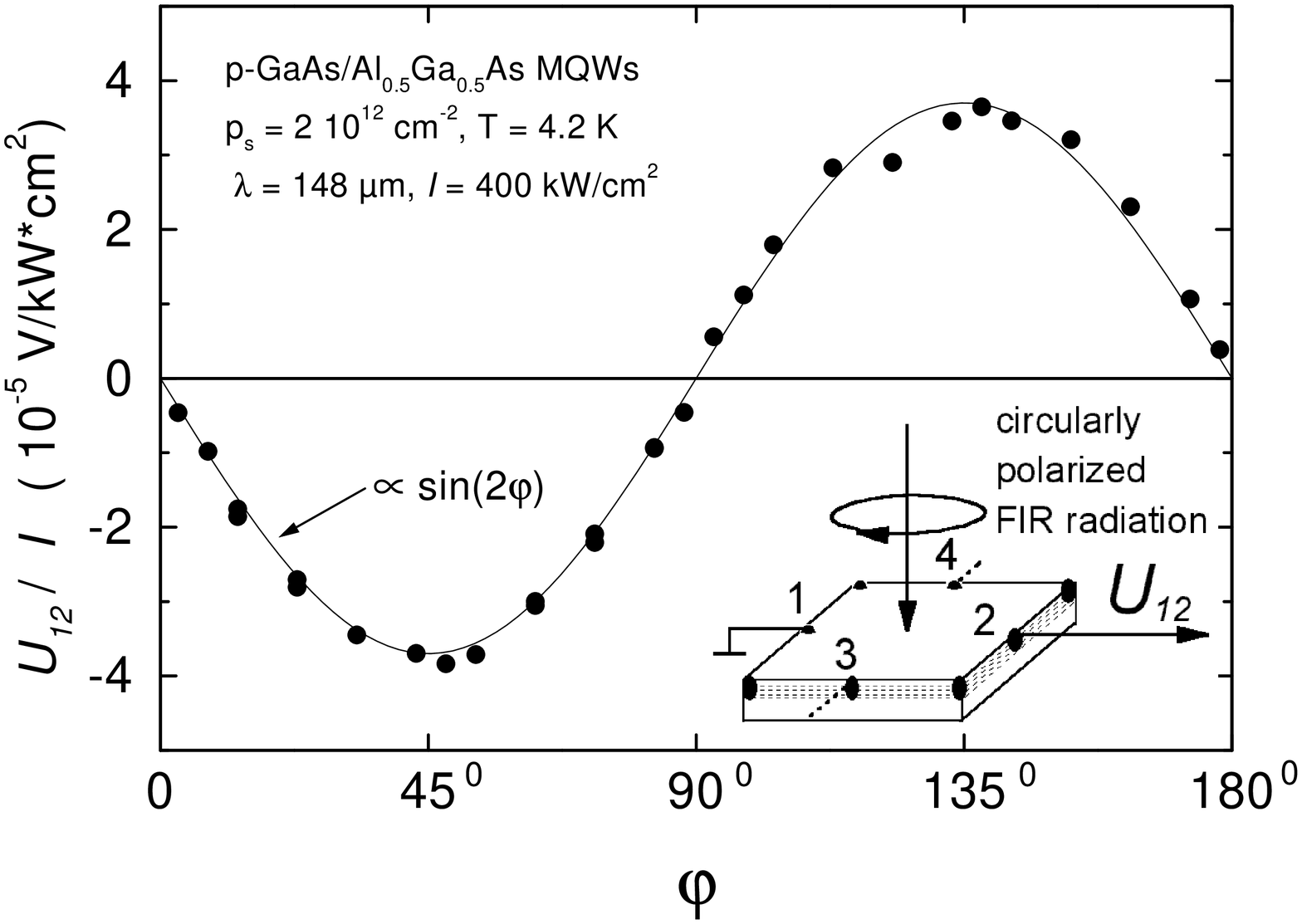}}
\caption{
Photogalvanic voltage signal $U_{12}$ picked up across the
contact pair 1-2 and normalized by the intensity $I$
as a function of phase angle $\varphi$.
The inset bottom-right shows the geometry of sample, contacts
and irradiation. The full line is fitted with one parameter after
$U_{12}\propto P_{circ}$ (see first line in
Eq.~3).
The orientation of opposite lying contact
pairs (1-2 and 3-4) with respect to crystallographic
axes is discussed in the text.
}
\end{figure}

\begin{figure}
   \centerline{\epsfxsize 10cm \epsfbox{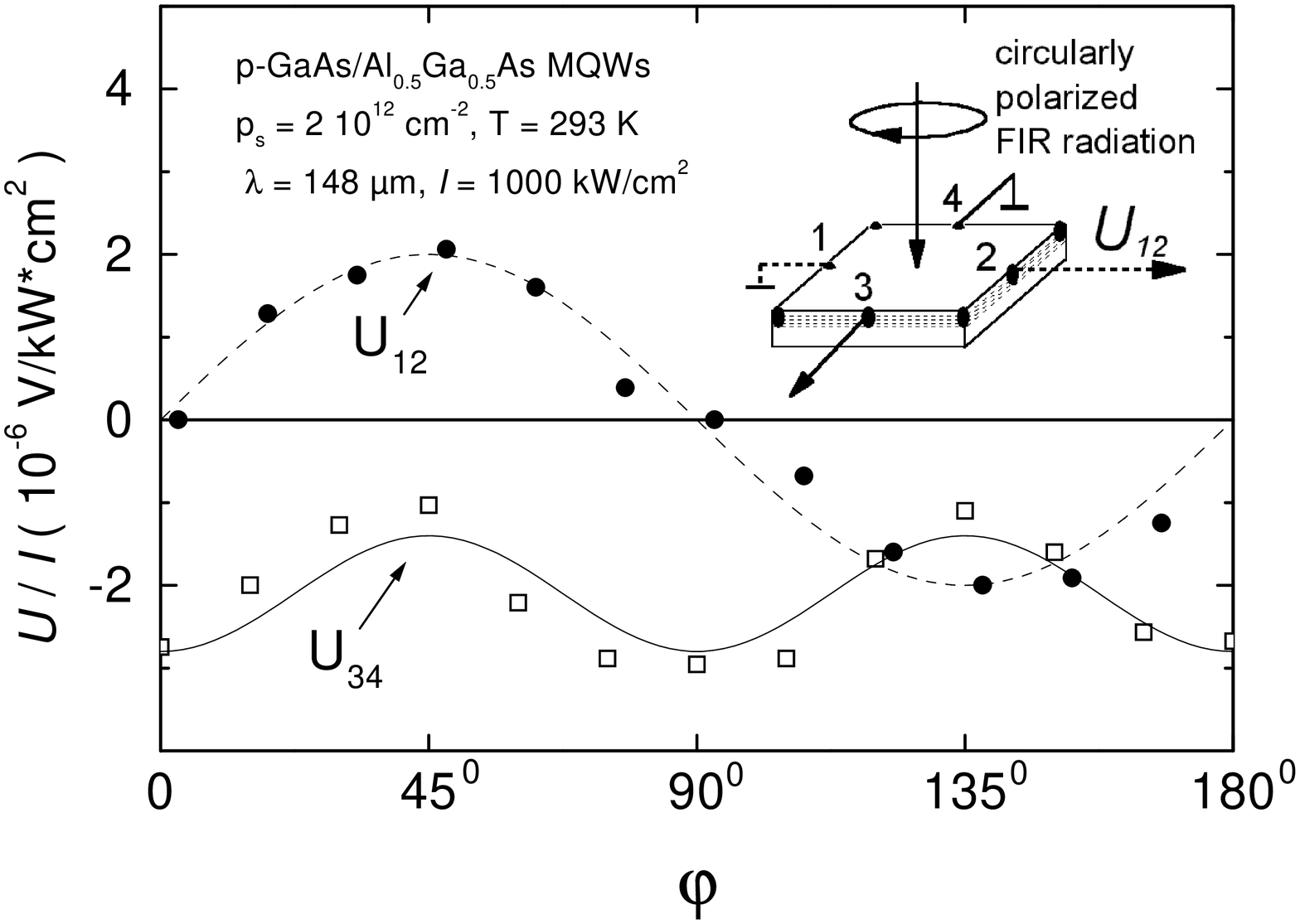}}
\caption{
Photogalvanic voltage signal $U$ (dots - $U_{12}$,
squares - $U_{34}$) normalized by the intensity
$I$ for the two different contact pairs with orthogonal
connecting lines (see inset) as
a function of phase angle $\varphi$. The broken line and the
full line are fitted after Eq.~ 3 as $U_{12} \propto \sin
2\varphi$ and $U_{34} \propto (\chi_+
+\chi_-\: \cos^2 2\varphi)$,
respectively.
}
\end{figure}

Using the other pair of contacts (contacts 3 \& 4 in the inset in
Fig.~2) no photogalvanic signal dependent on the sign of $P_{circ}$
could be detected. This is demonstrated in Fig.~2 where the voltage
$U_{34}$ is plotted as a function of the phase angle $\varphi$
together with $U_{12}$. The signal $U_{34}$ is periodic in
$\varphi$ with the period being one half of that of
$U_{12}$.
The value and the sign of $U_{34}$ for $\sigma_+$ are the same like that
for  $\sigma_-$
($\varphi = \pm 45^{\circ}$).
The measurements
of the photogalvanic effect along $\langle 100 \rangle$
crystallographic directions show a sum of the projections on this
direction of the photocurrents induced along [1$\bar{1}$0] and [110].

The photogalvanic current under study can be described by the
following phenomenological expression \cite{book}
\begin{equation}
j_{\lambda} = \chi_{\lambda \mu \nu} (E_{\mu} E^*_{\nu} + E_{\nu}
E^*_{\mu})/2 + \gamma_{\lambda \mu}\:i ({\bf E} \times {\bf E}^*
)_{\mu} \:,
\end{equation}
where {\bf E} is the complex amplitude of the electric field of the
electromagnetic wave and
\[
i\: ({\bf E} \times {\bf E}^*
) = P_{circ}\:E_0^2\:  \frac{{\bf q}}{q}\:,
\]
$E_0 = |{\bf E}|$, {\bf q} is the light wavevector. In a bulk crystal,
$\lambda = x,y,z$, while in a MQW structure grown along the $z$
direction, say $z \parallel [001]$, the index $\lambda$ runs only over
the in-plane axes $x \parallel [100], y \parallel [010]$ because the
barriers prevent carriers from moving along the $z$ axis and,
definitely, $j_z = 0$.

In bulk materials the photogalvanic effects under consideration arise
in homogeneous samples under homogeneous excitation due to an
asymmetry of the interaction of free carriers with photons, phonons,
static defects or other carriers in noncentrosymmetric media~\cite{book}.
The photocurrent given by the tensor $\bbox{\chi}$ describes
the so-called linear photogalvanic effect (LPGE) because it is usually
observed under linearly polarized optical excitation. The circular
photogalvanic effect (CPGE) described by the pseudotensor
$\bbox{\gamma}$ can be observed only under circularly polarized
optical excitation.

In an ideal (001)-grown MQW structure with the D$_{2d}$ point symmetry
or in a MQW structure with nonequivalent left and right interfaces
yet retaining their local symmetry C$_{2v}$, a generation of the
circular photocurrent {\it under normal incidence} is forbidden because
the pseudovector component $[{\bf E} \times {\bf E}^*]_z$ and the vector
components $j_x, j_y$ transform according to nonequivalent
irreducible representations of the group D$_{2d}$ or C$_{2v}$.
Thus in order to explain the experimental data the symmetry of the
structure should be reduced to the next possible point group which is
$C_s$ and contains only two elements: the identity transformation and
one mirror reflection, say in the plane (1$\bar{1}$0). In this case we
obtain under normal incidence
\begin{eqnarray} \label{1}
j_{x^{\prime}} &=& E_0^2\: [ \gamma P_{circ} + \chi_1 (e_{x^{\prime}}
e_{y^{\prime}}^* + e_{y^{\prime}} e_{x^{\prime}}^*)]\:
,\\
j_{y^{\prime}} &=& E_0^2\: [ \chi_2\: |e_{x^{\prime}}|^2 + \chi_3\:
|e_{y^{\prime}}|^2 ]\:, \nonumber
\end{eqnarray}
where {\bf e} is the polarization unit vector of the light and
$\gamma, \chi_1, \chi_2$ and $\chi_3$ are linearly independent
coefficients which can depend on the light frequency and temperature.
Instead of $x$ and $y$ we use the axes $x^{\prime} \parallel [1 \bar{1}
0], y^{\prime} \parallel [110]$. It follows then that the circular
photocurrent is induced along $x^{\prime}$, while the linear
photocurrent induced by the light linearly polarized along $x^{\prime}$
or $y^{\prime}$ flows in the $y^{\prime}$ direction. For the light
initially polarized along $x^{\prime}$ and transmitted through the
$\lambda/4$ plate we have
\begin{eqnarray} \label{pcirc}
j_{x^{\prime}} &=& \gamma E^2_0 \sin{2 \varphi} = \gamma E^2_0
P_{circ}\:, \\ j_{y^{\prime}} &=& E^2_0 ( \chi_+ + \chi_- \cos^2{2
\varphi} ) = E^2_0 ( \chi_2 - \chi_- P_{circ}^2 )\:, \nonumber
\end{eqnarray}
where $\chi_{\pm} = (\chi_2 \pm \chi_3 )/2$.
The CPGE described by $\gamma$ can be related to an electric current in
a system of free carriers with nonequilibrium spin-polarization
\cite{Ivchenko90}. The possible microscopic mechanism is the
spin-dependent scattering of optically oriented holes by asymmetric
interface imperfections. The experimentally observed change of sign of the
photogalvanic current upon heating the sample from 4.2~K to room
temperature may be caused by the change of scattering mechanism from
impurity scattering  to phonon assisted scattering. Note that, for ideal
zinc-blende-based MQW structures grown along the principal axis [001],
the CPGE is also possible but only under oblique incidence of irradiation.
\begin{figure}
   \centerline{\epsfxsize 10cm \epsfbox{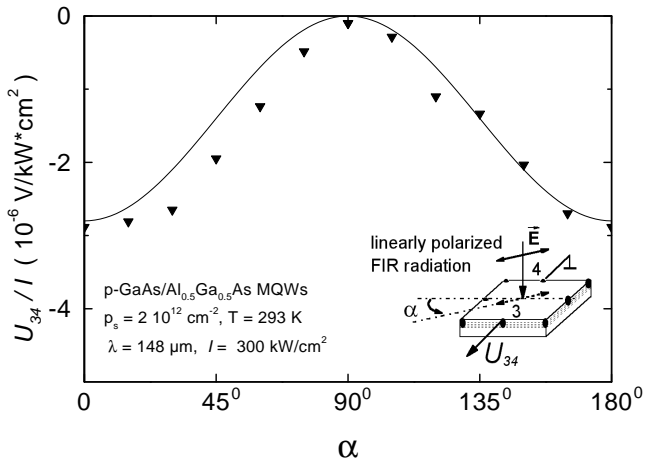}}
\caption{
Photogalvanic voltage signal $U_{34}$, in response to linear
polarized radiation, picked up across the
contact pair 3-4 and normalized by the intensity $I$
as a function of angle $\alpha$ between the electric field
vector and the connecting line of contact pair 1-2 (see
inset). Full line is fitted to $U_{34} \propto [ \chi_+
-\chi_-\: \cos2(\alpha+90^{\circ})]$ where the same parameters $\chi_+$
and $\chi_-$ have been used as for fitting $U_{34}$ in
Fig.~2. }
\end{figure}

Comparison of Eq.~(\ref{pcirc}) and Figs.~1,2 demonstrates a good
agreement between the theory and the experimental data. In order to
verify independently the validity of Eq.~(\ref{1}) for $j_{y^{\prime}}$
we excited the sample by linearly-polarized light and measured the
voltage $U_{34}$ as a function of the angle $\alpha$ between the plane
of linear polarization and the axis $x^{\prime}$. Note, that in this
set-up
$\alpha = 0$  is equivalent
to $\varphi = n\: 90^{\circ}$
($n$ is integer $0, \pm 1...$) in the set-up with the
$\lambda/4$ plate. Triangles in Fig.~3 present the measured dependence
$U_{34}(\alpha)$, the solid curve is calculated after
$U_{34} \propto[\chi_+ - \chi_- \cos{2 (\alpha+90^{\circ})}]$
which follows from the
second line of equation
(\ref{1}). To obtain the curve in Fig.~3 the same
parameters were used  as for fitting $U_{34}$ in
Fig.~2.

The low symmetry C$_{s}$ required to have nonzero $\gamma$ and
$\chi_{i}$ ($i=1,2,3$) gives an evidence for the in-plane
anisotropy showing that in the investigated MQWs (i) the [110] and
[1$\bar{1}$0] crystallographic
directions are different, and (ii) one of the reflection planes, (110)
either (1$\bar{1}$0), is removed as a symmetry element making the
corresponding normal axis to be a polar one. The first conclusion is
valid for both the $C_{2v}$ and $C_{s}$ point groups; the difference
between the [110] and [1$\bar{1}$0] axes has been studied by chemical
etching~\cite{Adachi83} of GaAs (001) surfaces and by an
observation of the in-plane anisotropy of the quantum Hall
effect~\cite{Lilly991} and anisotropic exchange splitting of the
excitonic levels (see~\cite{dzhioev} and references therein) in
periodic heterostructures. Moreover, (001) faces of GaAs show
trenches on the atomic level in one of the $\langle 110 \rangle$
directions~\cite{Newnham}. Obviously these trenches proceed into MQWs
forming interface islands, steps etc. and reducing their symmetry.
The conclusion (ii) is less obvious. It is worth to stress that, unlike
conventional optical and transport phenomena, the CPGE unambiguously
chooses between the $C_{2v}$ and $C_{s}$ symmetries. At present we can
only guess why the available series of MQWs is characterized by the
reduced symmetry. Photogalvanic studies of samples supplied from
different sources are needed to come to a final conclusion.

To summarize, it is shown that the optical orientation of free carriers
can be accompanied by their drift. Absorption of normally incident
circularly-polarized FIR radiation in gyrotropic media induces a
photocurrent which reverses the direction as the circular polarization
is changed from right- to left-hand orientation. These observations
reveal the reduced in-plane symmetry of GaAs-based (001)-grown MQWs
and can be used as a simple method to detect and characterize asymmetries
in heterostructures. The method is effective even at room
temperature. The theoretical considerations show that the CPGE can be
observed also in an ideal MQW of the symmetry D$_{2d}$ if the incident
radiation is obliquely incident. The experiments presented here
have been carried out in $p$-doped QWs but similar results are expected
also for  $n$-QWs.

Financial support by the DFG and the RFFI are gratefully
acknowledged. The authors would like to thank H.\,v.\,Philipsborn
for helpful discussions.

\end{document}